# Intrinsic Rashba Spin-Orbit Coupling in Staggered-Gyromagnetic Photonic Crystals

Yao-Ting Wang[1] and Wenlong Gao[2]

[1]*Department of Photonics, National Sun Yat-sen University, Kaohsiung, 80424, Taiwan*

[2]*Eastern Institute for Advanced Study, Eastern Institute of Technology, Ningbo, China*

**Abstract**

We report the realization of intrinsic Rashba spin-orbit coupling (SOC) in a two-dimensional photonic crystal composed of staggered-gyromagnetic cylinders in a modified honeycomb lattice. The system exhibits a Mexican-hat-like band structure and helical spin textures, which is the major characteristics of Rashba SOC. Through both full-wave simulations and *k-p* theory, we confirm the emergence of spin-split bands and vortex-like spin textures centered at the Brillouin zone. In addition, under oblique incidence, the Rashba band dispersion gives rise to concurrent negative and a positive refraction. These results establish a platform for exploring intrinsic Rashba photonics and spin-controlled wave transport in periodic systems.

The interplay between spin and momentum underpins a wide range of emergent phenomena in both condensed matter and photonic systems. Among all types of spin-orbit couplings (SOC), Rashba SOC represents a mechanism where inversion symmetry breaking and spin degrees of freedom give rise to spin-split band structures without the application of external magnetic fields. Rashba SOC manifests itself in the splitting of degenerate bands into two branches shifted laterally in momentum space, forming a pair of parabolic dispersions with opposite spin orientations. The corresponding spin texture exhibits a helical pattern in momentum space, where in-plane spin components wind around the Brillouin center. This spin-momentum locking results in two concentric iso-energy contours with opposite spin helicities, which is a defining signature of the Rashba SOC. Originally proposed to describe the spin-resolved electronic states in asymmetric quantum wells [1], the Rashba effect has since emerged as a cornerstone of spintronics and a fertile ground for realizing topological phases of matter [2]. Recent advances have further underscored relevance in spin orientation modulation [3] and electron interaction in Rashba materials [4].

In photonics, there has been a growing interest in the optical analogues of the Rashba effect. In such systems, the spin polarization of light locks to electromagnetic propagation direction. This photonic analogues of Rashba-like effect have been studied in a variety of systems, including tilted waveguide arrays [5], metasurfaces [6,7], Berry-phase engineered photonic crystal slabs [8], Dresselhaus-Rashba effect in perovskite cavities [9], and valley-photonic meta-structures [10] . These platforms typically employ artificial gauge fields or spatially varying geometric phases to emulate the Rashba Hamiltonian, producing spin-momentum locking or spin-dependent dispersion in an analogy to their electronic counterparts.

However, these demonstrations differ fundamentally from the intrinsic Rashba effect in periodic crystals despite their elegance. They rely on selective coupling of circularly polarized light to localized photonic modes, leading to spin-dependent photoluminescence in momentum space. These results mark crucial progress toward spin-controlled photonics but do not realize intrinsic Rashba feature of the photonic band structure. Similarly, another approach employs tilted waveguide arrays, where paraxial light evolution mimics Schrödinger-type dynamics. The setup demonstrates yields an effective Rashba Hamiltonian via synthetic gauge fields. Nonetheless, while the system exhibits distinct propagation modes linked with different spin degrees of freedom, the corresponding band structure presented does not display characteristics of Rashba SOC.

To date, the intrinsic Rashba SOC in photonic systems has long remained undiscovered. In this Letter, we demonstrate that Rashba SOC can be realized in photonic crystals with staggered gyromagnetic (SG) components and modified honeycomb lattice. Introducing time-reversal-symmetry breaking components is essential, for Rashba (or Dirac cone) like dispersions are incompatible with photonic time reversal symmetry operator [11], highlighting difference between photons and electrons. To this end, we derive an analytical expression of such system and introduce *k-p* theory to obtain the effective Hamiltonian near Γ point. This effective Hamiltonian links photonic wave equation with the standard methodology in quantum mechanics, thereby facilitating the visualization of the spin texture of every photonic band. The choice of gyromagnetic materials is yttrium iron garnet (YIG), which is an experimentally feasible matter due to its low loss property in microwave frequency range. Fig. 1a depicts the schematic of the two-dimensional SG photonic crystals with modified honeycomb lattice. The unit cell is extended to a hexagon that contains six cylinders, resulting in a four-fold Dirac point at Γ point because of the band folding of two Dirac degeneracies at K and K' points. Fig. 1b demonstrates how a four-fold Dirac point evolves. Starting with a regular honeycomb lattice, the implementation of SG constituents make the four-fold Dirac points split into a nodal line which is highlighted in blue circle. Then, the increase of distance from the center of unit cell lifts a complete gap separating two photonic band branches. Using both numerical simulation and $k$-$p$ theory, in the following we demonstrate that these bands can be directly mapped to Rashba SOC.

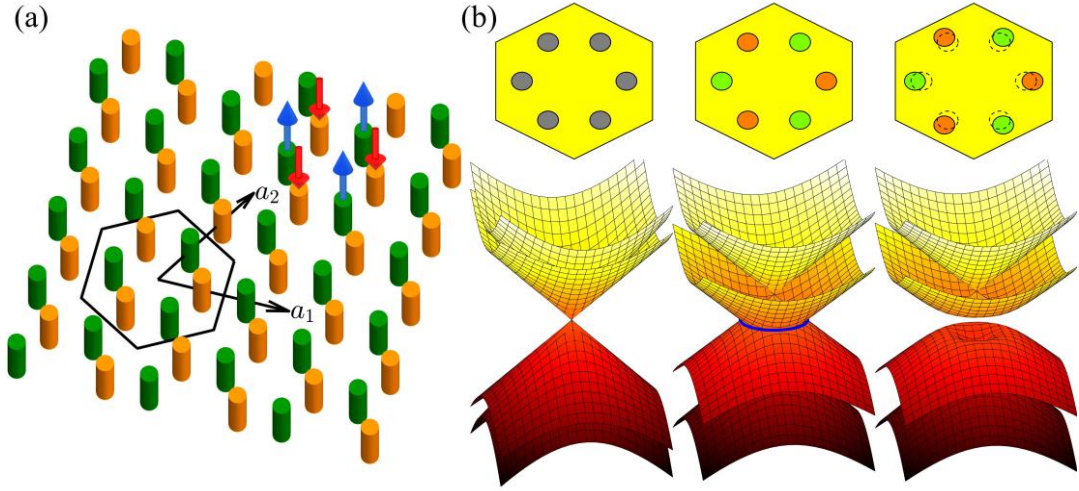

Fig. 1: Schematic of a hexagonal photonic crystal composed of two alternating types of cylinders. The green cylinders represent the gyromagnetic rods induced by a magnetic field in +z direction (blue arrows) whereas the orange ones represent the one induced by a magnetic field in -z direction (red arrows). The black hexagon denotes the unit cell, and the black arrows indicate the primitive lattice vectors $a_1$ and $a_2$. (b) Schematic of the photonic band evolutions as the material properties and structural geometry of the rods are modified. The top row shows the corresponding changes in the unit cell, while the bottom row presents the band structures, highlighting the transition from a Dirac cone to a nodal line, then to a Rashba band splitting.

The lattice constant of the proposed SG photonic crystal is denoted as $a$ and the radius of cylinders $r = 0.2R$, where $R$ is the distance of each cylinder to the center of unit cell. In a regular honeycomb lattice, $R = a/3$, whereas $R = 0.35a$ for a modified honeycomb lattice. The permittivity of YIG rods $\varepsilon = 15$ and the permeability tensor takes the form

$$\begin{bmatrix} \mu_a & -i\mu_g & 0 \\ i\mu_g & \mu_a & 0 \\ 0 & 0 & 1 \end{bmatrix}, \tag{1}$$

when an external uniform magnetic field applied in the out-of-plane direction. The elements of permeability tensor $\mu_a = 1 + \omega_0 \omega_m / (\omega_c^2 - \omega^2)$ and $\mu_g = \omega \omega_m / (\omega_c^2 - \omega^2)$, where the corresponding material parameters can be found in ref. [12]. When a 0.16T uniform magnetic field is applied, the approximate magnitude of $\mu_a = 14$ and $\mu_g = 12.4$ at frequency 4.281 GHz [13]. Here, the effect of material dispersion and loss are neglected so that the permeability tensor with real-valued $\mu_a$ and $\mu_g$ are both frequency-independent. Subsequently, we will show that the desired band structures are only slightly shifted due to dispersion. In the presence of magnetic anisotropy, the TE $(H_x, H_y, E_z)$ polarization gives rise to a 2D master equation where the conventional band theory is applicable to this system. By eliminating the magnetic fields from Maxwell's equations, one obtains

$$\nabla_\perp \cdot \xi(\mathbf{r}) \nabla_\perp E_z + i \left[ \nabla_\perp \times \eta(\mathbf{r}) \nabla_\perp E_z \right]_z + \left( \frac{\omega}{c} \right)^2 \varepsilon(\mathbf{r}) E_z = 0, \tag{2}$$

where the elements of inverse permeability tensors $\xi(\mathbf{r}) = \mu_a / (\mu_a^2 - \mu_g^2)$ and $\eta(\mathbf{r}) = \mu_g / (\mu_a^2 - \mu_g^2)$. The photonic band structure of this eigenvalue equation can be solved by plane wave expansion method or using numerical software such as COMSOL. Since the results from COMSOL and plane wave expansion are nearly identical, all the band structure presented in this Letter are all evaluated by COMSOL.

Fig. 2a-c illustrate the evolution of the TE band diagram for the realistic SG photonic crystals. For a regular honeycomb lattice without SG component, Fig. 2a contains a Γ-point four-fold Dirac degeneracy between the second and fifth bands at reduced frequency of 0.593. As mentioned previously, this degeneracy converts into a circular nodal line and two two-fold degeneracies when the SG material is employed, as shown in Fig. 2b. The nodal line is then lifted upon introducing a structural perturbation, thereby turning the entire four-band structure into a Rashba band diagram around reduced frequency of 0.52. In Fig. 2d, it is evident that the influence the material dispersion is insignificant within the desired frequency range. This validates that the

Rashba band splitting in Fig. 2d arises intrinsically from the special arrangement of gyro-magnetic materials, rather than dispersive properties, establishing the SG photonic crystal as a platform for realizing Rashba SOC in periodic media.

To construct a four-band reduced Hamiltonian near the Γ point, we use the Bloch wave function at $\mathbf{k}_0$ point as eigenbases because all the eigenstates can be obtained by numerical simulations. Since the Bloch wave function is known as $E_{z,n\mathbf{k}_0}(\mathbf{r}) = e^{i\mathbf{k}_0 \cdot \mathbf{r}} u_{n\mathbf{k}_0}(\mathbf{r})$ for eigenfrequencies $\omega_{n0}$, where $n$ denotes the band index, the

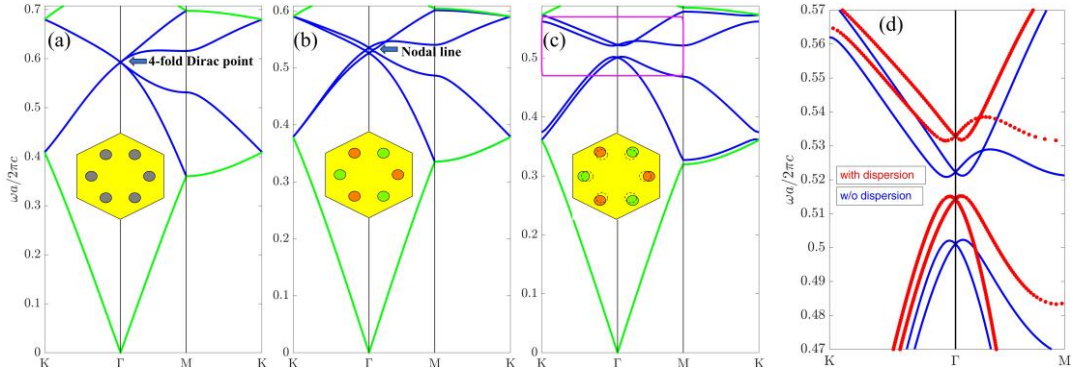

Fig. 2: Evolution of the photonic band structure under different perturbations. (a) The unperturbed structure supports a fourfold Dirac point at the Γ point. (b) Introducing staggered magnetic arrangement leads to the formation of a nodal line. (c) Further perturbation of the cylinders' positions results in the Rashba SOC. The insets in (a) to (c) illustrate the corresponding unit cell modifications. (d) Comparison of the zoomed-in band structure with (red) and without (blue) material dispersion, showing that the influence of dispersion is minor on the bandgap characteristics.

Bloch state at $\mathbf{k}$ point near $\mathbf{k}_0$ point takes the form [14]

$$E_{z,n\mathbf{k}}(\mathbf{r}) = e^{i\mathbf{k}\cdot\mathbf{r}} u_{n\mathbf{k}}(\mathbf{r}) = \sum_j A_{nj}(\mathbf{k}) \exp\left[i(\mathbf{k}-\mathbf{k}_0)\cdot\mathbf{r}\right] E_{z,j\mathbf{k}_0}. \qquad (3)$$

In here, the Bloch state $u_{n\mathbf{k}}(\mathbf{r})$ at $\mathbf{k}$ point has been expanded by the superposition of $u_{j\mathbf{k}_0}(\mathbf{r})$. Plugging (3) into (2) and employing orthonormal property of basis, we obtain the reduced Hamiltonian $H_{lj}$ given by

$$H_{lj} = -(\mathbf{k}-\mathbf{k}_0)\cdot\mathbf{p}_{lj} - i\left[(\mathbf{k}-\mathbf{k}_0)\times\mathbf{t}_{lj}\right]_z + (\mathbf{k}-\mathbf{k}_0)^2 q_{lj}, \qquad (4)$$

where the indices $l, j = 1 - 4$. In addition to the $p$ and $q$ coefficients that have been discussed in [15,16], the $t$ coefficient can be written as [Supplementary]

$$\mathbf{t}_{lj} = A_c^{-1} \int_{u.c.} E_{z,l\mathbf{k}_0}^*(\mathbf{r})(\nabla\eta) E_{z,j\mathbf{k}_0}(\mathbf{r}) d^2\mathbf{r}. \qquad (5)$$

with $A_c$ denoting the area of the unit cell. To construct this Hamiltonian, the calculation of (4) requires all Bloch states at $\mathbf{k}_0$ point. However, since the desired bands are significantly separated from other bands, we only need to consider the two pairs of degeneracies around Γ point to obtain the dispersion relation. The values of $\mathbf{p}_{lj}$, $\mathbf{t}_{lj}$, and $q_{lj}$ are presented in [17] and they form a 4-by-4 reduced Hamiltonian.

Notably, although the Eq. (4) does not explicitly appear as a direct sum of two conventional Rashba Hamiltonians owing to inter-band coupling among the four bands, it embodies a Rashba effect since the resulting band structure and spin texture are consistent with the nature of Rashba physics. To further validate whether the reduced Hamiltonian reflects underlying physics in our system, we compare the band structures calculated by Eq. (4) with the one calculated by COMSOL. These results have excellent

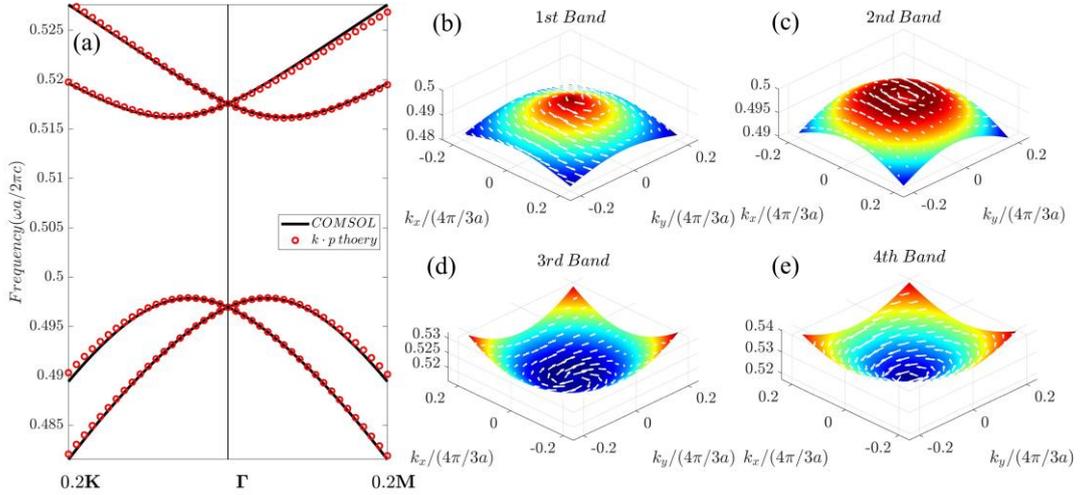

Fig. 3: Validation of the *k-p* band structure and the spin textures of eigenmode near the Dirac point. (a) Comparison of the band dispersion obtained from COMSOL simulations (black lines) and *k-p* theory (red circles) along the KΓ and ΓM directions. (b)-(e) Eigenfield patterns of the 1st to 4th Rashba bands, respectively, in the vicinity of the Γ point, showing the Rashba spin textures. The color represents the frequency distribution, while the white arrows indicate the in-plane spin components. The color map represents the spatial distribution of the real part of the electric field component.

agreement with the band diagram as shown in Fig. 3a. We then study the spin texture of each band. As stated in [18], the spin degrees of freedom in this six-rod structure is defined as $|\psi_n\rangle = [p_+, p_-, d_+, d_-]$, where $p_\pm = (p_x \pm ip_y)/\sqrt{2}$ and

$d_{\pm} = \left(d_{x^2-y^2} \pm i d_{xy}\right)/\sqrt{2}$. On this basis, the spin components can be evaluated through $s_i^{(n)} = \langle \psi_n | \mathbf{1} \otimes \mathbf{s}_i | \psi_n \rangle$, where $\mathbf{s}_i$ consists of three components of Pauli matrices. From the reduced Hamiltonian and its eigenstates $|\psi_n\rangle = [p_+, p_-, d_+, d_-]$, the SG photonic crystals give rise to unique spin textures that play a significant role in the propagation of light. As depicted in Fig. 3b-3d, the eigenfield patterns for the Rashba bands around the Γ-point clearly illustrate the spin textures associated with the photonic modes. The spin textures reveal how the pseudo-spin states of light evolve with the momentum of the photons in the SG photonic crystals. These spin textures describe that the photonic spin polarization, analogous to electron spin in the Rashba effect, is coupled to the momentum of the photons. As $k$ varies within the Brillouin zone, the direction of spin polarization of the photonic modes continuously rotates about the Brillouin center. This rotation of spin polarization is the main feature of Rashba SOC, which is a direct manifestation of the SOC in the system. Moreover, for different photonic bands near the Γ point, the spin textures are visualized as a vortex-like structure with opposite rotating directions. In Fig. 3b-3d, it is evident to see the two pairs of spin textures (1st-2nd band and 3rd-4th band) rotating in different orientation.

Fig. 4 illustrates the phenomenon of double refraction occurring in a SG photonic crystal slab illuminated by an obliquely incident electromagnetic wave at an angle of 20 degrees. The figure clearly demonstrates the splitting of the incident wave upon entry into the photonic crystal, generating two transmitted beams propagating at distinctly different angles. One refracted wave follows conventional positive refraction, with its wavevector bending toward the normal, consistent with predictions by the law of refraction for positive-index media. In contrast, the second refracted beam exhibits negative refraction, characterized by the wavevector emerging on the same side of the normal line as the incident wave. The directional arrows depicted in the figure explicitly demonstrate the energy flow of the reflected and refracted beams, highlighting the doubly refractive behavior. This anomalous double refraction is explained by analogy to the theory of resonant chiral media [19]. Since chiral resonance leads to Rashba-like band structure, the proposed SG structure produces similar conditions for spin states in photonic crystals. This dual refractive response is highlighted in the figure through directional arrows that indicate the group velocity and energy flow paths of each refracted beam.

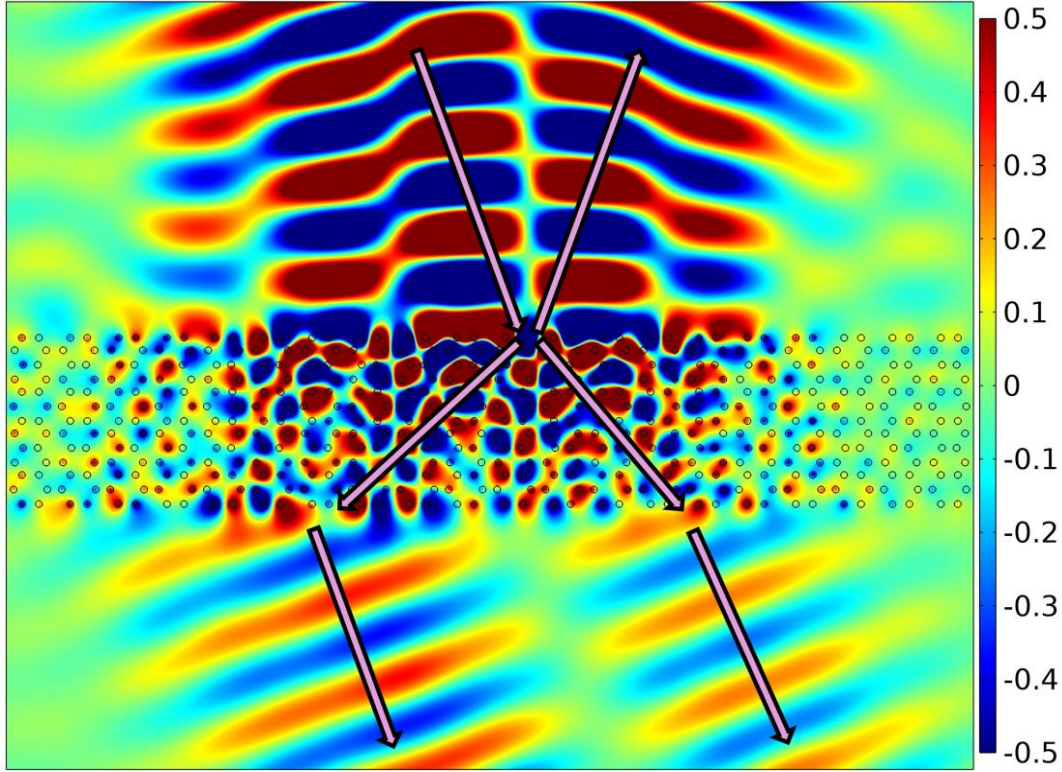

Fig. 4: Double refraction in a SG photonic crystal. A 20-degree incident wave interacts with the photonic crystal slab, resulting in two distinct refracted waves: one undergoing normal refraction and the other exhibiting negative refraction. The color map represents the field distribution, while arrows indicate the directions of wave propagation.

In conclusion, we have demonstrated that Rashba SOC can be effectively realized in SG photonic crystals with a modified honeycomb lattice structure. By employing YIG rods in a staggered arrangement in a hexagonal lattice, we achieve nontrivial magneto-optical properties which are equivalent to spin-orbit interactions analogous to the electronic Rashba SOC. Using both numerical simulations and analytical *k-p* theory, we achieve photonic band dispersion and validated these findings through excellent agreement with COMSOL simulations. Our analysis further reveals the formation of unique spin textures associated with each photonic mode near the Γ point. The spin polarization of the photonic modes continuously rotates around the Brillouin zone center, exhibiting a clear spin-dependent characteristic reminiscent of the Rashba effect. Moreover, the double refraction behavior demonstrates their capability to manipulate electromagnetic wave propagation uniquely, reinforcing their significance for advancing understanding in photonics. Future explorations into the topological and spin-dependent properties of these SG photonic crystals may reveal further intriguing physical phenomena.

**Acknowledgement**
YTW acknowledges financial support from the National Science and Technology Council under Grant number: NSTC 113-2112-M-110-007-MY3. WG acknowledges by the National Natural Science Foundation of China under Grant number: 12474309.